\documentclass{aastex}

\def\go{\mathrel{\raise.3ex\hbox{$>$}\mkern-14mu\lower0.6ex\hbox{$\sim$}}}
\def\lo{\mathrel{\raise.3ex\hbox{$<$}\mkern-14mu\lower0.6ex\hbox{$\sim$}}}
\slugcomment{PASP Millenium Essay}
\shorttitle{The Future of Gravitational Lensing}
\shortauthors{Blandford}
\begin{document}
\title{The Future of Gravitational Optics}
\author{R. D. Blandford}
\affil{130-33 Caltech, Pasadena, CA 91125}
\begin{abstract}
In this speculative, millenial essay, I try to anticipate what sort 
of novel gravitational optics investigations might be observed, after it  
becomes possible to map and monitor roughly a trillion sources 
(of which a billion may be usefully variable)  comprehensively 
throughout electromagnetic and other spectra over the whole sky. 
Existing techniques suffice to produce three dimensional maps of the 
dark matter distribution of the accessible universe, to
explore black hole spacetimes and to magnify images of the first 
luminous sources, terrestrial planets and compact objects.
\end{abstract}
\section{The Past Millenium}
This has been a good millenium for the study of 
gravitational optics. The foundations of the subject 
were set by the scholastics, Grosseteste, Bacon and Theodoric in 
the 12th and 13th 
centuries with their studies of rainbows and mirages. Four centuries
later, Newton used Galilean kinematics to formulate his laws of 
motion and gravitation and develop his
corpuscular theory of light using the laws of refraction, 
as previously elucidated by Snell and Descartes. Meanwhile, 
Fermat put forward the principle of 
least time, which made the connection between optics and
dynamics which was cemented though the remarkable 
insights of Euler, Lagrange and 
Hamilton culminating in the development of 
quantum mechanics in the twentieth century. 
Independent of these formal developments, Michell, Laplace,
Soldner and others quantified the magnitude of 
the deflection of light by gravitational fields
according to Newton's theories, although their 
answer had to be doubled to allow for space curvature after Einstein 
perfected his general theory of relativity.

Observational progress had to wait until the last quarter of the 
twentieth century. (It is impressive how much that we now 
observe in detail was anticipated, notably by Zwicky, Refsdal, 
Gunn and Paczy\'nski.) Following the first report of a 
strong (multiple-imaging)
gravitational lens by \citet{wal79}
we have subsequently found cluster arcs \citep{sou87}, weak
lensing in clusters \citep{tys90}, galaxies \citep{bra96}
and the field \citep{van00} 
as well as microlensing \citep{irw89,alc93}. These observations 
have had a big impact on cosmology 
and have already been used to estimate the size and shape of the universe, 
as tracers of dark matter and as natural telescopes.

It is therefore with some optimism that the gravitational optics community
can face the present millenium. What can we expect as we observe
this spendid optical bench from our (presumably unprivileged) 
vantage point in spacetime? In this essay, I would like to take the 
long view and imagine that one day, the whole universe within 
our past light cone will be comprehensively monitored throughout the
electromagnetic and other spectra and that the exabytes 
of data that are produced will be automatically stored 
and searched.
\section{Cosmological Sources}
From the deep imaging of the Hubble Deep Fields, it is apparent that Eddington
was correct and that there
are $\sim10^{11}$ OIR sources on the sky to $\sim30^m$, roughly 
2'' apart on average. It is likely 
(and I shall assume) that most of these are protogalactic sub-components 
with $z\sim1-3$, though 
this has not yet been demonstrated. We will surely count these sources
to fainter magnitudes and image overlapping galaxies so as to account for
the brightness of the sky in essentially all bands in which the 
universe is transparent to $\sim$ milliarcsecond resolution so as to
provide a background of $\sim10^{12}$ 
matchable sources for performing gravitational lensing investigations.
(The fluctuating cosmic microwave background provides a similar source 
but it is unlikely to have observable structure on angular scales 
that will be seen by any but the largest lenses.)

Roughly $10^7$ galaxies have bright enough nuclei to be called 
quasars and there are $\sim10^8$ X-ray sources and $\sim10^{10}$ compact 
radio sources \citep{kel01}. In round numbers, I estimate that 
there are $\sim10^9$ continuously varying, compact sources
that will be monitored.
Explosive sources are also interesting. Whole sky
searches to $z\sim3$ should yield a supernova rate $\sim3$~s$^{-1}$
unless they are mostly obscured \citep{por00}. Gamma ray bursts (GRBs) will
surely be traced to the faint end of their luminosity function at cosmological
distances suggesting rates at least as large as $\sim10^{-3}$~s$^{-1}$
\citep{bla99}. (GRBs may
also offer the opportunity to observe lensing effects asssociated
with neutrinos and gravitational radiation.) For these purposes,
afterglows are similar to supernovae. 
\section{Galaxy Lenses}
A good way to think about the type of images that can form is 
to imagine a pencil of rays propagating backwards in time from Earth
(in the scholastically-approved manner), 
past lensing galaxies and other masses and focusing on caustic surfaces.
Sources located on rays between the first and second foci will
form inverted images at Earth and be accompanied by at least two 
positive parity images formed by different rays (unless there is 
obscuration) \citep{sch92}. The comoving volume occupied
by these putative sources, which I will call the lensing volume,
is a measure of the number of observable instances
of multiple imaging to be found. Double and quad image configurations
are usually seen. The latter are rarer per source but are 
over-represented in existing observations because they are more 
easily recognized and because they are strongly magnified.
Most lenses are elliptical, on account of their greater 
central concentration and the denser environments in which they are
found \citep{koc99}.  The influence of additional mass 
along the line of sight can also influence the imaging in a manner 
that will be quantifiable on an individual basis. Most galaxy lenses
turn out to have $0.5\lo z\lo1$ and there are $\sim10^9$ 
of them. If we imagine a typical lens with Einstein
radius $\sim1''$, then the lensing volume to $z\sim3$ is 
roughly 1 Mpc$^3$ per lens.  The total lensing 
volume containing strongly lensed sources 
is $\sim1$~Gpc$^3$,   
roughly one thousandth of the total comoving volume out
to $z\sim3$.  This is consistent
with the observed lensing probability per distant 
source $\sim0.001$ and leads to an estimate of roughly one lensed source 
per lensing galaxy on average.

These considerations lead to an estimate of $\sim10^9$ observable 
instances of gravitational lensing, though some of these will be hard to see 
against the light of the lensing galaxies.
Point sources that lie close enough to caustics will create
pairs of images that are linearly magnified along their lines of 
separation. It should be possible to identify them efficiently 
on spectral grounds. The lensing volume per lens with magnification  
$>A$ is $\sim A^{-2}$~Mpc$^3$ and the most magnified sources 
of this type
should have $A\sim3\times10^4$, provided that they have detectable 
structure on a scale $\sim3A^{-1}$~kpc in order to remain 
unresolved by the gravitational lens.
If one in a thousand of these sources is a variable AGN then the 
greatest intrinsic magnification in a variable source
should be $A\sim10^3$. Multiply-imaged
supernova (GRBs) should occur every few minutes (weeks) and 
the brightest supernova (GRB) 
over a millenium should have $A\sim10^4, (10^2)$.

More interesting are the cusp lines which run along the caustic surfaces. 
Sources within and close to these cusps form bright, colinear triples
in which the central image should have a 
flux equal to the sum of the fluxes from the other two images.
The lensing volume where the weakest of the 
three images has magnification $\go A$, can be estimated to be 
$\sim A^{-5/2}$~Mpc$^3$. The brightest galaxy cusps should have 
magnifications $A\sim10^3$, if the sources are compact enough.

The next most prevalent catastrophe is known as the hyperbolic 
umbilic and this has a three dimensional caustic
surface \citep{pet01}. There are two of these associated
with every elliptical lens possessing a finite core. 
They can form four image patterns
looking like normal quads but all four images lie 
on one side of the lens center. (The order in
which these images varies differs from that found in regular 
quads and so they are distinguishable.) They are likely to be superposed
on the lens galaxy, and so are best sought using radio VLBI.
Hyperbolic quads are very sensitive to the 
shape of the galaxy potential but are strongly magnified. 
If, as a calculation of a plausible galaxy model suggests
is the case, the relevant lensing volume for forming them is 
$\sim0.01$~Mpc$^3$ per lens, then there should
be $\sim10^4$ hyperbolic quad radio sources on the sky.

A second four image, three dimensional catastrophe is the swallowtail. 
This can form when there is a close binary pair of galaxies. Four, 
roughly colinear images with alternating parity are formed although
there are should be at least one other image in the vicinity.
Groups and clusters are the best places to look for swallowtails
and there could be 
thousands of examples which, if analyzed, would provide excellent
probes of the galaxy potential on large scales. 

The final three dimensional, generic catastrophe is the elliptical umbilic. 
In its most likely manifestation, this involves three negative parity
images surrounding a positive parity image although there should be 
at least two additional, observable images in the vicinity. They are most
likely to be located inside a triangle formed by three lensing galaxies
with appropriate separations \citep{rus01}.
Although triple galaxies are less common than double galaxies, the 
lensing volume
per galaxy within which a source will form a recognizable elliptic umbilic
is larger and these may outnumber swallowtails.
\section{Clusters and Compact Groups}
Clusters and groups influence lensing in two separate ways. They 
can magnify sources already lensed by galaxies, making them more 
common and easier to observe and they can act as lenses in their own right.
Of course, all clusters and groups can multiple-image through their brightest
galaxies. Although the groups probably dominate the total lensing volume
the cluster lenses will continue to be of special interest
because it is possible to map them with more confidence
and precision. In round numbers, there are $\sim10^5$ 
multiply-imaging clusters 
each with a lensing volume $\sim100$~Mpc$^3$. This implies that the probability
of a distant source being multiply imaged 
by a cluster is $\sim10^{-5}$. Cluster-induced caustics are capable of 
very large magnification \citep{bla96}
and future, space-borne interferometers should
be capable of resolving individual bright stars, magnified a million times as
they cross the caustic curve.

Our observations need not be passive. Typical sources are likely
move up to a pc per millenium. This is enough for us to observe secular
changes in the most strongly magnified sources and envisage more elaborate
experiments. For example, suppose that we launch an array of 
robotic telescopes and use three of them to measure the velocity of a caustic
sheet (formed by a bright source) as it passes Earth, a fourth 
telescope could be made to ``surf'' the wave and observe the 
source with considerable magnification for a long time. 
\section{Weak Lensing Tomography}
Weak lensing studies are, necessarily, statistical in character. However,
when $\sim10^{10}$ image shapes are measured over the sky, then it should be 
possible use them to map the dark matter distribution, 
assuming that we can determine their redshifts statistically \citep{tys00}.
A typical, individual potential fluctuation on a nonlinear scale 
length $\sim30$~Mpc.
creates a shear of $\sim10^{-3}$. There will be roughly 
100 of these regions along a line of sight whose effects add
stochastically giving a typical total shear $\sim10^{-2}$.
Using the $\sim10^6$ sources behind each of these regions, we should
be able to measure the depths of their individual potentials with 
$O(1)$ accuracy. 
Measurements on larger scales should have slightly larger signal to noise.
(A much greater number of source shapes is necessary 
to map typical regions of space as opposed to the much smaller number
that already suffices to map large fluctuations like rich 
clusters.) The density of usable sources 
is unlikely to be high enough to permit the mapping of 
individual galaxies, except in those cases where there are relatively 
bright source galaxies along the optic axis. However statistical 
studies of galaxies as a function of type, redshift, size etc 
are quite feasible.
\section{Black Holes and Neutron Stars}
The best prospects for exploring the spacetime around a black hole,
lie with the three million solar mass hole at our Galactic center. 
It will be possible to track faint stars as they approach the 
black hole and predict passages as close as $\sim100$ gravitational
radii. This should permit quantitative tests of the Kerr metric.
Advanced pulsar search techniques using giant radio telescopes
should lead to discovery of thousands of binary pulsars in nearby galaxies.
Some of these will be observed from within $\sim10'$ of the orbital plane 
and fairly accurate tests of relativity will become possible. 
\section{Stars}
Our Galaxy has over $\sim10^{11}$ stars, most of which will
be typed and monitored. Giant, space-borne interferometers 
should be capable of measuring their proper motions and their distances
will be known quite accurately on the basis of parallax, spectroscopy
and photometry. As the microlensing optical depth 
is $\sim10^{-6-7}$, stellar  
events should be initiated roughly every hour and
should be predictable. Neutron star lenses, which will be mostly unanticipated,
should be found weekly and black hole events, perhaps monthly. Double star
lenses are particularly interesting because they form intricate
caustics and there is a good chance that microlensing will provide
the primary search technique for distant planets. Furthermore,
terrestrial planets might be studied and their images
deconvolved during a caustic crossing if they trail their star and are 
magnified a thousandfold for about an hour under the most favorable conditions.
\section{When will all this happen?}
The Lyapunov time for contemporary astrophysics is
so short that I cannot predict how long it will take to map and monitor
the whole sky to faint flux levels. However, as I am supposed
to be in the business of prediction, let me guess that 
someone reading this essay, this year, will witness attempts
to accomplish most of what I have briefly
discussed plus much more that I lack the imagination to
anticipate. Unfortunately, I do not know if this prediction is one 
in the field of astronomy or medicine!
\acknowledgments
I acknowledge support under NSF grant AST-9900866 and thank Leon Koopmans
for discussions and the editors for their patience.

\end{document}